\documentclass[aps,prl,twocolumn,superscriptaddress,showpacs]{revtex4}

\usepackage{graphicx}
\usepackage{epsfig}

\begin{document}
\title{Observation of a mixed pairing state in Ru microdomains embedded in Sr$_{2}$RuO$_{4}$}

\author{Z. Long}
\affiliation{Department of Physics, The Pennsylvania State University, University Park, PA 16802}

\author{C. Andreou}
\affiliation{Department of Physics, The Pennsylvania State University, University Park, PA 16802}

\author{Z.Q. Mao}
\affiliation{Department of Physics, Tulane University, New Orleans, LA 70118}

\author{H. Yaguchi}
\affiliation{Department of Physics, Kyoto University, Kyoto 606-8502, Japan}

\author{Y. Maeno}
\affiliation{Department of Physics, Kyoto University, Kyoto 606-8502, Japan}
\affiliation{International Innovation Center, Kyoto University, Kyoto 606-8501, Japan}

\author{Y. Liu}
\email{liu@phys.psu.edu}
\affiliation{Department of Physics, The Pennsylvania State University, University Park, PA 16802}

\date{\today}

\begin{abstract}
We report first measurements on tunneling into the {\it interior} of a Ru microdomain embedded in bulk chiral $p$-wave superconductor Sr$_{2}$RuO$_{4}$. The junctions were prepared by pressing a pure In wire onto the ${\it{ab}}$ face of a Ru-containing Sr$_{2}$RuO$_{4}$ single crystal. Below the superconducting transition temperature ($T_{c}$) of Sr$_{2}$RuO$_{4}$ (= 1.5~K), we observed a zero-bias conductance peak (ZBCP) associated with the proximity-induced $p$-wave superconductivity. However, below 0.44 K, roughly the $T_{c}$ of bulk Ru (= 0.5~K), an unexpected Ru gap and a broad ZBCP superimposed with subgap features were found to emerge, with unusual magnetic field dependences. We argue that the new features resulted from a mixed pairing state featuring intrinsic ${\it{s}}$- and ${\it{p}}$-wave condensates in the interior of a Ru microdomain embedded in Sr$_{2}$RuO$_{4}$.
\end{abstract}






\maketitle

Superconductors are divided into two categories featuring an even-parity, spin-singlet and odd-parity, spin-triplet pairing symmetry, respectively. The inversion symmetry requires that the pairing symmetry of a bulk superconductor be of either an even or odd parity. However, in the absence of such symmetry, as in crystals with no inversion center \cite{Bauer} or near a superconductor-normal metal interface \cite{Edelstein}, mixed pairing states are allowed by symmetry, and realizable under suitable conditions. 

Pairing symmetry of a superconductor is determined primarily by effective interactions. In the presence of a rotational symmetry, the interaction can be decomposed into channels with different orbital angular momentum, $l$. The superconducting energy gap, $\Delta_l$, is given by $\Delta_l = 2\varepsilon_{l} exp(-2/N_0V_l)$, where $\varepsilon_{l}$ is a characteristic energy, $N_{0}$ is the density of states (DOS) at the Fermi energy, and $V_l$ is interaction in the $l$ channel \cite{Mineev}. States of $l$ = 0, 2 ... then correspond to ${\it{s}}$-, ${\it{d}}$-wave ...(even-parity) and those of $l$ = 1, 3 ... to  ${\it{p}}$-, ${\it{f}}$-wave ... (odd-parity) pairings. Electrons will usually pair in the channel with the largest $V_l$, corresponding to the highest superconducting transition temperature ($T_c$), to attain a largest condensation energy, even though other $V_l$'s may be substantial.

Chiral ${\it{p}}$-wave superconductivity has been established in Sr$_2$RuO$_4$ by many experiments \cite{Mackenzie2003} including recent phase sensitive measurements \cite{Nelson}. The eutectic Ru-Sr$_2$RuO$_4$ system of pure Ru microdomains (a few $\mu$m in size) embedded in bulk Sr$_2$RuO$_4$ \cite{Maeno1998}, which is of fundamental interest \cite{Sigrist2001} because of the enhanced $T_c$ as high as 3 K, has emerged as a useful system for the study of unconventional pairings \cite{Mao, Kawamura}. Bulk Ru is an ${\it{s}}$-wave superconductor, as evidenced by the insensitivity of superconductivity to impurities \cite{William}. The intriguing questions are how the intrinsic pairing in $s$-wave channel competes for stability with the proximity-induced ${\it{p}}$-wave pairing below the intrinsic $T_{c}$ of Ru and whether intrinsic pairing in $p$-wave channel is possible in the interior of a Ru microdomain embedded in Sr$_2$RuO$_4$. 

To answer these questions, we prepared junctions of In/Ru/Sr$_2$RuO$_4$ on Sr$_2$RuO$_4$ single crystals grown by floating-zone technique \cite{Mackenzie2003}. These crystals are essentially pure crystals containing only few scattered Ru microdomains. A Sr$_2$RuO$_4$ crystal was first cleaved along its ${\it{ab}}$ plane. Then a freshly cut, high-purity (5N) In wire of a 0.25 mm diameter was pressed onto the ${\it{ab}}$ face of the crystal without any artificial tunnel barrier. The current-voltage ($I$-$V$) characteristics of the junctions were measured with a standard 4-terminal, DC technique in a $^3$He and dilution refrigerator with a base temperature of 300 and 20~mK, respectively. The tunneling conductance d$I$/d$V$ were obtained by taking the numeric derivatives of the $I$-$V$ curve.

Even though no structural characterization of the junctions was carried out, the totality of the tunneling and transport data allows us to divide the junctions into three types. For the first type, the tunneling is dominated by a contact between In and the nonsuperconducting surface layer \cite{liuLT} of pure Sr$_{2}$RuO$_{4}$, yielding a tunnel spectrum well described by the Bardeen-Cooper-Schiffer (BCS) formula \cite{Tinkham} with an In gap of 0.5 meV. For the second type, the tunneling is dominated by a contact between In and the 3-K phase, the interface region between Ru and Sr$_2$RuO$_4$, showing an In gap around 0.5~meV and a zero-bias conductance peak (ZBCP) resulted from Andreev surface bound states (ASBSs), similar to those seen previously \cite{Mao, Kawamura}. For the third type of junctions, the focus of the present work, the tunneling is dominated by a contact between In and the interior of a Ru microdomain (Inset of Fig. 1a). 

\begin{figure}
\includegraphics{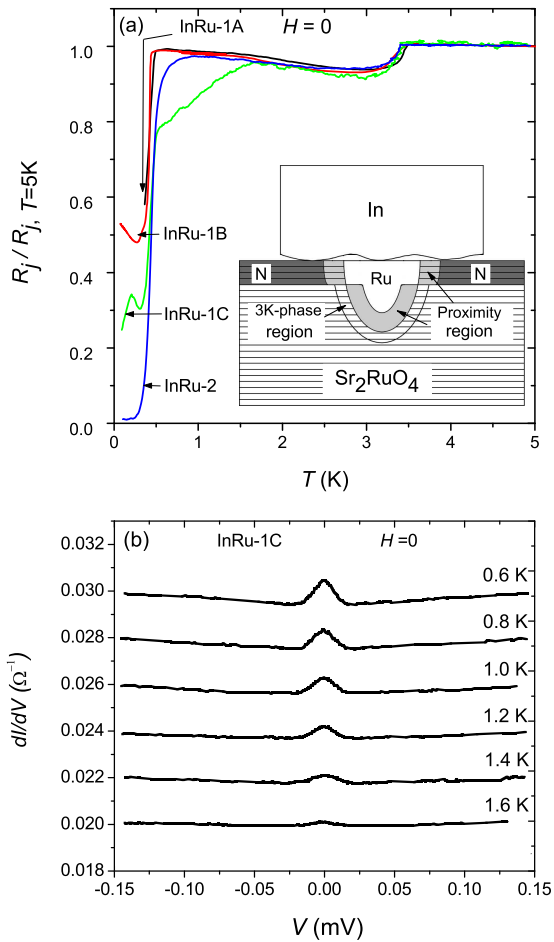}
\caption{(color online) (a) $R_{j}(T)/R_{j}(5K)$ vs $T$ for two In/Ru/Sr$_{2}$RuO$_{4}$ junctions. InRu-1 was measured in three separate cooling-downs - in a $^3$He refrigerator with a $c$-axis field (InRu-1A) and a dilution refrigerator with a $c$-axis (InRu-1B) and an in-plane (InRu-1C) field. Values for $R_{j}(5K)$ are 47.5, 51.2 and 60.6 $\Omega$ for InRu-1A, B, and C and 12.5 $\Omega$ for InRu-2. Inset: Schematic of the third type of In/Ru/Sr$_{2}$RuO$_{4}$ junctions. The nonsuperconducting surface layer (N) is indicated; (b) d$I$/d$V$ vs $V$ for the InRu-1C for $T \geq 0.5~K$ in the zero field. Curves except the bottom one (1.6 K) are shifted upward by 0.01 ~$\Omega^{-1}$ for clarity.} 
\end{figure}

Measurements on the temperature-dependence of the junction resistance, $R_{j} (T)$, revealed a negative d$R_{j} (T)$/d$T$ above 3.4~K (Fig. 1a), suggesting the existence of a barrier between In and Ru, probably due to the formation of InO$_x$ on In surface, resulting in a superconductor-insulator-superconductor ($SIS'$) junction. $R_{j} (T)$ showss a drop at the $T_c$ of In, 3.4 K, for all samples, and a barely visible kink around 1.5~K, the $T_c$ of bulk Sr$_2$RuO$_4$ for all samples except InRu-1C. For the latter, a much more pronounced kink was found at a slightly higher temperature (1.7~K). We believe that the shift is due to the development of additional contacts between In and Ru close to, but not at the 3-K phase region after two thermal cyclings. No feature was found between 1.7 to 3~K, suggesting that the In contact is away from the 3-K phase in all samples. As the temperature is lowered, $R_{j} (T)$ exhibits an abrupt drop around the $T_c$ of bulk Ru, 0.5~K. For InRu-1B and C, a non-monotonic $R_{j} (T)$, to be discussed below, is seen below 0.5 K. 

Tunnel spectra $T \geq 0.5~K$ are shown in Fig. 1b for InRu-1C. Similar spectra were obtained for other samples. A rather small ZBCP was found as high as 1.6~K, above the $T_c$ of bulk Sr$_2$RuO$_4$. Such ZBCP is very similar to those found previously in tunneling into the 3-K phase region where Andreev surface bound states (ASBS) are formed on the sample surface of the 3-K phase \cite{Mao}. These ASBSs are formed in this case because of an exponentially small, proximity-induced $p$-wave order parameter in the interior of the Ru microdomain. As the sample was cooled to below 0.44~K, slightly lower than the $T_c$ of bulk Ru (0.5~K), new features emerge in the tunnel spectrum, including a sharp drop in the tunneling conductance at a well-define gap edge, $\Delta_{Ru}$ (=e$V_{Ru}$), and a ZBCP (Fig. 2a). Again almost identical features were found for all other samples. 

\begin{figure}
\includegraphics{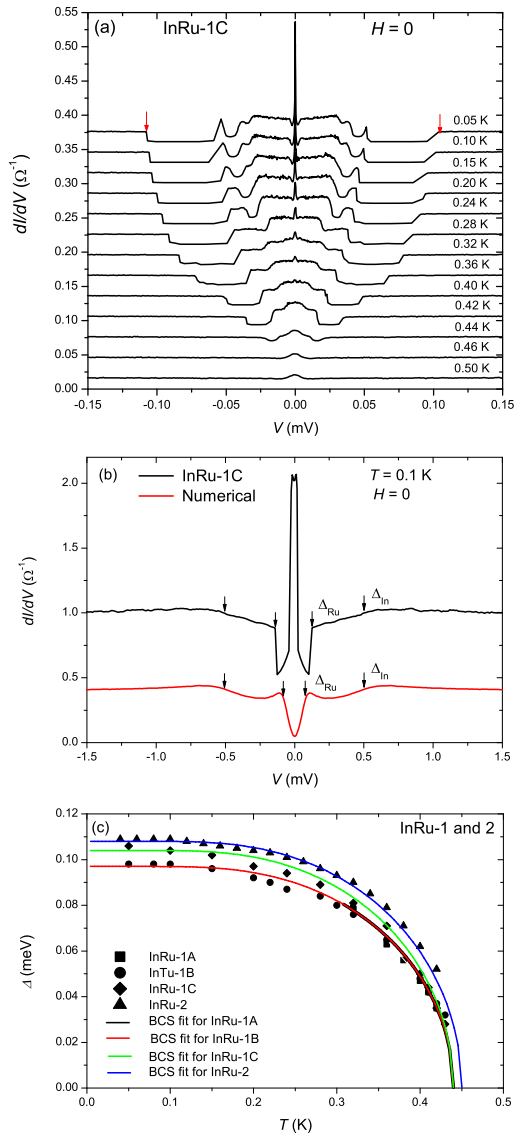}
\caption{(a) d$I$/d$V$ $\it{vs} $ $V$ for the InRu-1C for T $\leq $ 0.5 K in zero field. Curves except the bottom one are shifted upward by 0.03 $\Omega^{-1} $ for clarity. The red arrows indicate $V_{Ru}$; (b) Normalized d$I$/d$V$ $\it{vs} $ $V$ for the InRu-1C in zero field at 0.1 K (upper curve with $\Delta_{In (Ru)} $ = 0.50 (0.10) meV). The lower curve shows a numerical simulation of a conventional In/Ru junction at $T$ = 0.1 K (shifted by 0.6 $\Omega^{-1} $), employing an effective DOS of the form, $N_{eff}(E) = r (N(E) - 1) + 1)$, where $N (E)$ is the broadened BCS DOS (Eq. 2) and r accounts for the junction leakage ($N_{eff}(E) = 1$ at high energies). The parameters used in the simulation are: $\Delta_{In (Ru)} $ = 0.50 (0.072) meV, $\Gamma_{In (Ru)} $ = 0.20 (0.03) meV, and $r_{In (Ru)} $ = 0.15 (0.50); (c) Un-normalized $\Delta_{Ru} $ $\it{vs} $ $T$ for all samples. Solid lines are BCS fits to $\Delta_{Ru}\it{vs} T$, yielding  $\Delta_{Ru} $ = 0.097, 0.097, 0.104, and 0.108 meV, and $T_c$ = 0.44, 0.44, 0.44 and 0.45 K for InRu-1A, B, C, and InRu-2, respectively.}
\end{figure}

For such $SIS'$ junction, gap features would normally be expected at $\Delta_{In}$ $\pm $ $\Delta_{Ru}$. However, the spectrum can change when broadening is introduced. Starting from the normalized $I-V$ characteristic for a $SIS'$ junction, 

$$I(V) = \int^{+\infty}_{-\infty}N(E)N'(E+V)[f(E)-f(E+eV)]dE \eqno{(1)}$$

where $E$ is the energy, $f(E)$ is the Fermi function, $N(E)$ and $N'(E)$ are the normalized, broadened DOS for $S$ and $S'$ of the form 

$$N(E) = Re[\frac{\mid E \mid+i\Gamma}{\sqrt{(\mid E \mid+i\Gamma)^2-\Delta^2}}] \eqno{(2)}$$

where $\Gamma$ is the broadening parameter reflecting, for example, a finite quasiparticle lifetime \cite{Dynes}, we can obtain the normalized tunnel conductance $G_j$ ($=dI/dV$). Assuming that In and Ru are ${\it{s}}$-wave superconductors, we indeed obtain gap features at $\Delta_{Ru}/e$ and $\Delta_{In}/e$ in our simulation, as shown in Fig. 2b. Interestingly, similar results were obtained for high-$T_c$ junctions\cite{Hu1998, Chesca}.  Identifying $\Delta_{Ru}$ as the Ru gap is further supported by the excellent fit of $\Delta_{Ru}(T)$ to the BCS formula (Fig. 2c). 

Another important observation is that the ZBCP evolves into a continuous plateau superimposed by distinct peaks below 0.44 K. The conductance values of the plateau were found to be larger than twice of the normal-state conductance, characteristic of unconventional pairing, suggesting that $p$-wave pairing continues to be present at low temperatures. The satellite peak, found around $\Delta_{Ru}$/2 in the ZBCP just below 0.44 K and seen to split into two at the lowest temperatures, should correspond to subgap resonant states. The sharp central peak in the spectrum, however, must come from the Josephson coupling between In and Ru. If the order parameter direction of proximity-induced $p$-wave state is the same as that in the bulk, as expected, Josephson coupling between the $s$-wave In and proximity-induced $p$-wave Ru is forbidden by the selection rule \cite{Jin2000}. Therefore the observed Josephson coupling may indicate the emergence of a new superconducting component. 

The development of a gap edge close to the $T_c$ of bulk Ru suggests that the intrinsic $s$-wave pairing of Ru may be this new superconducting component, suggesting futher pairings in both $l=0$ and $1$ channels in the interior of a Ru microdomain below 0.44 K. The interesting question is whether the $p$-wave pairing can be intrinsic, rather than proximity-induced. Such simultaneous pairings in both ${\it{s}}$- and ${\it{p}}$-wave channels in a Ru microdomain is not forbidden by symmetry. The typical size of a Ru microdomain is a few $\mu$m. The superconducting coherence length of bulk Ru, $\xi_{0,Ru}$, in the clean limit, is $\approx $ 4$\mu$m based on $\xi_{0,Ru}=\hbar v_{F}/\pi\Delta_{0,Ru}$, where $v_{F}$ is the Fermi velocity and $\Delta_{0,Ru}$ is energy gap ($v_{F}$ = 1.5 $\times$ 10$^6$ m/s and $\Delta_{0,Ru}$ = 0.072 meV). An inversion operation on such a Ru microdomain does not constrain the wavefunction symmetry of the Cooper pair as its size is comparable with or larger than the sample size. Microscopically, $V_l$ usually decreases rapidly with increasing $l$, making $V_1$ the second largest among $V_l$s. A condensate in the $l=1$ channel in the Ru microdomain, with a smaller condensation energy than that of a $l=0$ condensate, could be stabilized by energy saving from minimizing the gradient of the order parameter between the Ru microdomain and the 3-K phase of Sr$_2$RuO$_4$.

This scenario is consistent with the quantitative result on $\Delta_{0,Ru}$ (= 0.1~meV), which is larger than the BCS value, 0.072~meV, estimated from $T_c$ = 0.44 K. The ratio, $2\Delta_0/k_BT_c$, has an unexpectedly high value ($\approx$ 5.5), typical of a strong-coupling superconductor, which is difficult to accept because of the low $T_c$. In fact, from $T_c = 1.14T_{\theta}exp[-1/N(0)V]$ \cite{Tinkham} and $T_{\theta,Ru}$ $\approx$ 404 K, we find that $[N(0)V]_{Ru}$ = 0.147, comparable with that of a protypical weak-coupling superconductor, Al, with $[N(0)V]_{Al} \approx$ 0.165. On the other hand, if we assume a mixed pairing state with the form \cite{Tsuei}:
$$\Delta_{Ru} = \Delta_{s}+i{\Delta_{p}}\eqno{(3)}$$
where $\Delta_{s}=i\sigma_{2}\Delta_{0}$ is the ${\it{s}}$-wave and $\Delta_{p}=i\sigma_{2}\vec{\sigma}\cdot\vec{d}$ the ${\it{p}}$-wave gap function, with $\vec{d}$ the vector order parameter and $\vec{\sigma}=(\sigma_{1}, \sigma_{2}, \sigma_{3})$ Pauli matrices \cite{Mineev}. If $\Delta_{p}$ belongs to a one-dimensional representation and is real with a magnitude similar to that of $\Delta_{s}$, $\mid \Delta_{s} \mid \approx$ $\mid \Delta_{p} \mid \approx$ 0.072 meV, we find that $\mid \Delta \mid = \sqrt{\mid \Delta_s \mid^2+\mid\ \Delta_p \mid^2}$ = 0.102 meV, which agrees well with experiment. This scenario also provides a plausible explanation for the non-monotonic behavior in $R_{j} (T)$, seen also in InRu-2 under a magnetic field. $R_{j} (T)$ of an $SIS'$ junction measured above its critical current, $I_c$, follows the temperature dependence of $I_c$($T$). The Josephson couplings between In and the two individual superconducting components in Ru could be of different signs, leading to a non-monotonic $I_c$($T$) and $R_{j} (T)$, as seen previously \cite{Jin1999}.

The magnetic field dependence of tunnel spectrum provides further support to the mixed pairing picture. Assuming that the field dependence of $\Delta_{s}$ is essentially that in the bulk, a type I superconductor with a $H_c$ of 69 Oe, $\Delta_{s}$ will be fully suppressed at 69 Oe. As shown in Fig. 3, with a field applied along the $ab$ plane, $\Delta_{Ru}$ is seen to be suppressed rapidly as the field increase to about 70 Oe and more slowly for larger field. Within the mixed pairing picture, $\Delta_{s}$ vanishes at this field, leading to $\Delta_{Ru}$ = $\Delta_{p}$. Experimentally, $\Delta_{Ru}$ = 0.07 meV at 70 Oe, agreeing with the mixed pairing picture.

\begin{figure}
\includegraphics{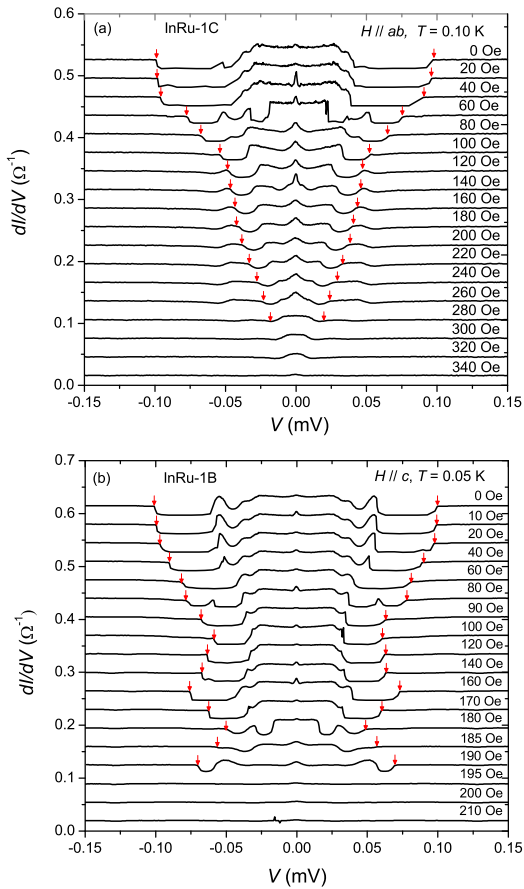}
\caption{d$I$/d$V$ vs $V$ for InRu-1 ($T$ = 0.1 K) with the field applied along the $ab$ plane (a) and the $c$ axis (b). Curves except the bottom ones are shifted upward by 0.035~$\Omega^{-1} $ for clarity.  The red arrows track the field variation of $\Delta_{Ru}$. For field applied along the $c$ axis, the spectrum depends on whether the sample is zero-field- or field-cooled. Data presented here were obtained after the sample was field-cooled from 5 K.}
\end{figure}

An intriguing evolution of tunneling spectrum was observed when the field was applied along the $c$ axis. As seen in Fig. 3b, $\Delta_{Ru}$ shows a non-monotonic behavior - with the increasing field, the gap and ZBCP features seen at lower fields were found to re-emerge at higher fields. Qualitatively similar behavior was observed in InRu-1A and in InRu-2. Since Sr$_{2}$RuO$_{4}$ is a chiral $p$-wave superconductor \cite{Mackenzie2003}, chiral currents flowing in the $ab$ plane are expected on the surface of the Ru microdomain. When the interior of the Ru microdomain becomes superconducting, a chiral order parameter given by Eq. 3 will lead to its own chiral current. When a magnetic field is applied along the $c$ axis, a finite flux will thread through the in-plane chiral current loop. We believe that the flux will modulate the chiral current just as the flux modulates the supercurrent in Little-Parks effect \cite{Tinkham}, leading to $T_c$ and gap modulations. This process will minimize (maximize) the gap edge ($\Delta_{Ru}$) for certain fields, as seen experimentally. Such modulation will not occur when the field is applied along the in-plane direction.

\begin{acknowledgments}
We would like to acknowledge useful discussions with Profs. D.~Agterberg and M.~Sigrist. The work is supportedby DOE under DE-FG02-04ER46159 at Penn State, by Louisiana Board of Regent under LEQSF (2003-06)-RD-A-26 at Tulane, and by JSPS and NEXT of Japan at Kyoto. Z.Q. Mao is a Cottrell scholar.
\end{acknowledgments}


\end{document}